
\magnification=\magstep1
\nopagenumbers
\def\Th#1{\Theta_{#1}}

\def\ps#1{\Xi_{ #1}}
\def\dd{\hbox{d}}
\def\tr{\hbox{tr}}
\font\tif=cmr10 scaled\magstep3
\font\tii=cmti10 scaled\magstep3
\def\big{1}
\def\th{2}
\def\bz{3}
\def\m{4}
\def\w{6}
\def\al{\alpha}
\baselineskip=18truept
\rightline{iassns-hep-92-47}
\vskip -4truept
\rightline{6/92}
\vfil
\centerline{\tif Non-analyticity in the
large \tii N \tif renormalization group }
\vfil
\centerline{Vipul Periwal\footnote{${}^\dagger$}{vipul@guinness.ias.edu}}
\bigskip
\centerline{The Institute for Advanced Study}
\centerline{Princeton, New Jersey 08540-4920}
\vfil
\par\noindent The flow of the action induced by changing
$N$ is computed in large $N$ matrix models.
It is shown that the change in the action is non-analytic.
This non-analyticity appears at the origin of the space of
matrices if the action is even.
\vfil\eject
Given analogies between the double-scaling solutions${[\big]}$ of matrix
models and critical phenomena, it is natural to consider $N$ as a
kind of cutoff in the string theory, associated with the matrix
model by means of the Feynman diagram expansion of the
matrix model${[\th]}$.
One can motivate this interpretation of $N$
quite precisely.  Double-scaling gives the following form for the
universal part of the free energy of the matrix model, equivalently the
partition function to all orders in the genus expansion of the
associated string theory:
$$ \eqalign{F &\sim \ \sum_{\chi} b_{ \chi} (g_{
c}-g)^{\chi\al} N^{ \chi}\cr
&= \sum_{\chi}b_{ \chi} N^{\chi}
g_{ c}^{\chi\al}
\sum_k \left({g\over g_{ c}}\right)^k
{\Gamma(\chi\al+1)\over\Gamma(k+1)\Gamma(\chi\al-k+1)}\cr
&= \sum_k \left({g\over g_{ c}}\right)^k
\sum_\chi b_{ \chi}\left(Ng_{ c}^{\al}\right)^{ \chi}
{\Gamma(\chi\al+1)\over\Gamma(k+1)\Gamma(\chi\al-k+1)}.\cr
}$$
Now, if we look at the series for large values of $k,$ we find
$$
{\Gamma(\chi\al+1)\over\Gamma(k+1)\Gamma(\chi\al-k+1)}
\sim\  (-)^k {k^{-\chi\al}\over\Gamma(-\chi\al)},$$
implying that the coefficients of $(-g/g_{ c})^{ k}$ for large  $k$
are functions of $A\equiv k/N^{1/\al}g_{ c}.$  Since $k$ is
the number of plaquettes on the triangulated surface, and $\al>0,$  it is
natural to interpret $A$ as the renormalized area, with
$N,k\rightarrow\infty$ as the continuum limit.  In other words,
one interprets $N^{-{1/\alpha}}$ as the area of a plaquette.
It therefore is of interest to compute
the renormalization group flow induced by varying $N.$

Having given this motivation, it is necessary to point out that
the above manipulation is only of heuristic value.  The asymptotic
behaviour of the coefficients is only valid for $k$ much larger than
$-\chi\al,$ so one cannot deduce properites of the fixed renormalized
area partition functions by looking at the large order behaviour of the
free energy.  It would be too much of a digression to say more about
such an approach in the present paper.

Br\'ezin and Zinn-Justin${[\bz]}$ recently considered a
large $N$ renormalization
group approach to matrix models, with different motivations in mind.
They obtained results that suggested that qualitative properties of
the double-scaling solutions could be reproduced by such
considerations.

The precise problem is the following:
Let $\ps{N+1}$ be an $(N+1)\times (N+1)$ Hermitian matrix, and let
$\ps{N}$ be related to $\ps{N+1}$ as follows:
$$\ps{N+1} \equiv \left(\matrix{\ps{N}&\psi\cr
\psi^{* \rm T}&\xi\cr}\right),$$
where $\psi$ is a vector with $N$ complex entries and
$\xi$ is a real number. Let $V$ be a polynomial\footnote{${}^*$}{$V$ %
will be an even polynomial throughout this paper for simplicity.} of the
form $\sum_{i=1}^{ M} a_ix^{2i}/2i.$  Then, with
$\Omega_{N}\equiv \hbox{vol.}U_{ N}$ (in the adjoint representation),
we wish to find $\delta V$ such that
$${\cal Z}_{ N+1}\equiv
\int {\dd\ps{N+1}\over\Omega_{N+1}} \exp\left[-(N+1)\tr
V(\ps{N+1})\right] =
\int {\dd\ps{N}\over\Omega_{N}} \exp\left[-N\tr
(V+\delta V)(\ps{N})\right] .$$

I shall show that the evolution of $V$ under changes of $N$
induces non-analytic changes in $V$ around the
origin of the matrix integration domain.  The resulting
matrix model may not possess a simple surface interpretation,
and more importantly, that the large $N$ renormalization group is not
defined on the space of polynomials in the basic operators
$\hat\Th{n}\equiv\hbox{tr}~\Xi^n/N.$

I use a measure that agrees explicitly with standard matrix model results.
My results are physically in accord with what one expects in
random matrix physics${[\m]}.$
In an eigenvalue picture, integrating out one eigenvalue redefines the
potential felt by the other eigenvalues.  In particular, suppose we
look at the effective potential felt by the remaining eigenvalues
near the origin matrices.  The non-analyticity arises from
the following physics:  to leading order in $N,$ the evaluation
of the contribution to $\delta V$ due to the eliminated eigenvalue
is given by this eigenvalue attempting to be as far from the
origin as possible, consistent with the fact that it is confined by the
potential, $V.$  There are two such positions available to this eigenvalue.
However, only
one of these saddlepoints contributes to leading order in $N.$  It is
precisely the discrete change in the dominant saddlepoint value for
an  infinitesimal change in the matrix $\ps{N}$ that
leads to the non-analyticity in the effective potential.
Of course, as with all non-analytic behaviour associated with
an `infinite volume' (here, the limit of large $N$), the non-analyticity
is ameliorated when `finite-size' effects (here, $1/N$ corrections) are
accounted for.

While the models considered are simple enough to allow
analytic solution, the result obtained here
is concrete evidence for the subtlety of the
large $N$ limit in general.  In particular,
the extraction of physics at finite $N,$ including $1/N$ corrections,
for theories with fields transforming in the adjoint representation,
may be more subtle than one might expect na\"\i vely.

I now turn to the solution of the problem stated above.
The first observation is that the $U_{ N+1}$
symmetry can be broken to $U_{ N},$ with the
eliminated generators used to set $\psi=0.$  This produces a
Jacobian factor in the measure so
$${\cal Z}_{ N+1} =
\int {\dd\ps{N}\over\Omega_{N}} \exp\left[-(N+1)\tr
V(\ps{N})\right] \int \dd\xi \exp\left[-(N+1)V(\xi)\right]
\det\left(\xi-\ps{N}\right)^2.$$
Thus
$$N\tr\delta V = \tr V - \hbox{ln}~\int\dd\xi\exp\left[-(N+1)
V(\xi)\right]\det(\xi-\ps N)^2.$$
It remains therefore to evaluate the integral over $\xi.$
This integral can be evaluated by saddlepoints since the
determinant is that of an $N$-dimensional matrix, and $N+1$
multiplies $V$ in the measure.  This evaluation is
non-perturbative as far as $V$ is concerned for we need make no
assumptions about the coefficients in $V.$


\def\ar{5}
We want to figure out the changes in the polynomial $V$ when we
integrate out some of the degrees of freedom.  We may assume
that $\hat\Th n $ are small in the following,
since the coefficients of a polynomial can be
obtained from its behaviour at the origin.
The saddlepoint value of $\xi_s$ satisfies
$$\xi_sV'(\xi_s) =
{2}{1\over N+1}\hbox{tr}~{1\over \xi_s-\ps{N}}=
{2}{N\over N+1}\left[
1 +{\hat\Th{1}\over \xi_s}
+{\hat\Th{2}\over \xi_s^2} +\dots\right] .\eqno(1)$$
A very similar equation appears in double-scaling solutions of
vector models[\ar] as well---this should not come as a surprise since
we have essentially integrated out a vector here.
We can solve this equation  for $\xi_s=\xi_s(a_i,\hat\Th{n}).$
If we are interested in perturbation theory, in other words in the
surfaces provided by the Feynman diagram expansion,
it is of interest to solve eq.~1 when $a_i, i >1$ are assumed small.
To leading order in $N$ we can
set $N/(N+1)\approx 1$ in eq.~1,   since the error is of
the same order as corrections to the saddlepoint value of the
integral.

Eq.~1 has two solutions with the same assumptions
as above.  Since we consider $V$ even,
these are related by $\xi_s\rightarrow -\xi_s.$  Then the change in $V$ is
$$\eqalign{\tr\delta V(\ps{N}) &=
{1\over N} \left[\tr V(\ps{N}) - \hbox{ln}\left\{
\exp\left[-(N+1)V(\xi_s)\right]\left(\det(\xi_s-\ps{N})^2 +
\det(\xi_s+\ps{N})^2\right)\right\}\right]\cr
&= {1\over N}\tr V(\ps{N})
+ {N+1\over N}V(\xi_s) \cr
&\phantom{= {1\over N}\tr V(\ps{N})}
-\hbox{ln}\left\{
\exp \hbox{tr~ln}~(\xi_s-\ps{N})^2 + \exp\hbox{tr~ln}~(\xi_s+\ps{N})^2
\right\}^{1/N}\cr &= V(\xi_s)+
{1\over N} \left[\tr V(\ps{N}) + V(\xi_s)- \hbox{max}\left(
\hbox{tr~ln}(\xi_s-\ps{N})^2,
\hbox{tr~ln}(\xi_s+\ps{N})^2\right)\right],\cr}$$
up to terms higher order in $1/N.$
The maximum function arises as the limit
$\lim_{n\rightarrow\infty}(a^n+b^n)^{1/n}= \max(a,b).$
The monotonicity of the exponential function implies
$\exp~\hbox{max}(a,b)=\hbox{max}(\hbox{e}^a,\hbox{e}^b).$
We  have thus derived non-analytic behaviour in $\delta V.$
Terms involving $\hat\Th{n}$ were neglected in eq.~1, consistent with
the fact that I am investigating
the change in the potential at $\ps{N}\approx 0.$

Very close to the origin, up to constants,
$$\tr\delta V \approx \ -{2 }\left|{\hat\Th1\over\xi_s}\right|,\eqno(2)$$
which is an even function, as expected.  Na\"\i vely one could write
$$\tr \delta V =  \left(1+ {1\over N}\right)V(\xi_s)
-2 \hbox{ln}|\xi_s|
+ {1\over N} \tr V(\ps{N})
-2\left\{\left|
{\hat\Th1\over\xi_s}+
{\hat\Th3\over3\xi_s^3}+\dots\right|
- {\hat\Th2\over2\xi_s^2}
- {\hat\Th4\over4\xi_s^2}- \dots \right\}.$$
This expression neglects the presence of $\hat\Th{i}$ in
eq.~1, hence of terms that are nonlinear in $\hat\Th{i}$ in $\delta V.$
The linear term displayed in eq.~2 giving the
behaviour of $\delta V$ near the origin is unchanged by such corrections.

Since the non-analyticity arises in the Gaussian theory as well, one might
attempt to define a relative renormalization, by subtracting the
non-analyticity found in a Gaussian theory.  However, it is easy to see
that this does not remove the non-analytic behaviour.  Consider
$V= ax^2/2 + bx^4/4,$ then
$$\xi_s =\pm \sqrt{{2\over a}}\left(1- {b\over a^2}+
\dots\right) .$$
If we ignore the non-analyticity, we derive
$$\eqalign{\delta a &= {2a\over N}  \left(1+{b\over a^2} - {2b^2\over
a^4} +\dots \right),\cr
\delta b &= {b\over N}\left({a^2\over 2b} + 3 - {2b\over a^2} +\dots
\right).\cr}$$
If we rescale $a$ to its original value, we find
$$\tilde b = b +{1\over N} \left(a^2 -b - {6b^2\over a^2} +\dots\right)
+\dots .$$
Thus, if we now subtract the renormalization of $b$
that would occur even in a free theory, we find the result of
Ref.~\bz~(eq. 43).  It therefore becomes clear how to make contact
with the results of Ref.~\bz.  Note that the
saddlepoint evaluation of the $\xi$ integral is valid for $b>-a^2/4,$
which implies that the critical value $b_c = -a^2/6$ is within the
validity of the evaluation.

It should be stressed that in the theory of the renormalization group,
approximate recursion relations are extremely important, so the
renormalization found by Br\'ezin and Zinn-Justin may well be
the appropriate approach to the large $N$ renormalization group.
However, it is important to understand
the nature of the approximation.  It is in this context that the
calculation given in this paper is of interest.  Furthermore, it was
shown here that there are terms induced in the potential that do not
allow a perturbative surface interpretation.  However, if we
started with a potential that was not even, we would obtain
non-analytic behaviour at other points in the integration domain.
The flow of even
a Gaussian integral is rather subtle when the large $N$ symmetry group's
volume is taken into account.  It would be fascinating if this
was related to the properties of the topological `critical' point
found by Witten${[\w]}.$

\bigskip
I am grateful to O. Lechtenfeld, R. Myers and C. Nappi for helpful
conversations, and to J. Zinn-Justin for an e-mail communication.
This work was supported by D.O.E. grant DE-FG02-90ER40542.

\medskip

\centerline{References}
\medskip
\item{\big .} E. Br\'ezin and V. Kazakov, {\sl Phys. Lett.} {\bf 236B}
(1990) 144; M. Douglas and S. Shenker, {\sl Nucl. Phys.} {\bf
B335} (1990) 635; D. Gross and A. Migdal, {\sl Phys. Rev. Lett.} {\bf
64} (1990) 127
\item{\th .} G. 't Hooft, {\sl Nucl. Phys.} {\bf B72} (1974) 461
\item{\bz .} E. Br\'ezin and J. Zinn-Justin, E.N.S.--Saclay preprint
LPTENS 92/19/SPhT/92-064 (1992)
\item{\m .} M.L. Mehta, {\it Random matrices}, Academic Press (New York,
1967)
\item{\ar .} A. Anderson, R.C. Myers and V. Periwal, {\sl Phys. Lett.}
{\bf 254B} (1991) 89, {\sl Nucl. Phys.} {\bf B360} (1991) 463;
S. Nishigaki and T. Yoneya, {\sl Nucl. Phys.}
{\bf B348} (1991) 787; P. di Vecchia, M. Kato and N. Ohta,
{\sl Nucl. Phys.} {\bf B357} (1991) 495
\item{\w .} E. Witten, {\sl Nucl. Phys.} {\bf B340} (1990) 281
\end